 \documentclass[preprint2]{aastex}

\newcommand{\mathsym}[1]{{}}
\newcommand{\unicode}[1]{{}}
 \usepackage{graphics}
 \usepackage{graphicx}
 \usepackage{epsfig}

\usepackage{amssymb}
\usepackage{amsmath}
\usepackage{aas_macros}

\shorttitle{Normalization of Hamiltonian \dots with P-R drag in non-resonance case}
\shortauthors{Ram Kishor et al.}

\begin{document}


\title{Normalization of Hamiltonian and nonlinear stability of triangular equilibrium points in the
photogravitational restricted three body problem with P-R drag in non-resonance case}


 \author{Ram Kishor}
 \affil{Central University of Rajasthan, NH-8, Bandarsindari, Kishangarh, Ajmer-305801, Rajasthan, India}
 \email{kishor.ram888@gmail.com; kishor\_math@curaj.ac.in}
  \author{M. Xavier Rames Raj}
  \affil{Applied Mathematics Division, Vikram Sarabhai Space
 Centre, Thiruvananthapuram-695022, India}
 \email{xavierjamesraj@gmail.com}
  \author{Bhola Ishwar}
  \affil{Department of Mathematics, B. R. A. Bihar
 University, Muzaffarpur-842001, India}
 \email{ishwar\_bhola@hotmail.com}

\begin{abstract}
Normal forms  of Hamiltonian are very important to analyze the nonlinear stability of a
dynamical system in the vicinity of invariant objects. This paper presents the normalization of Hamiltonian and the analysis of nonlinear stability of triangular equilibrium points in non-resonance case, in the photogravitational restricted three body problem
under the influence of radiation pressures and P-R drags of the radiating primaries. The Hamiltonian of the system is normalized up to fourth order through Lie transform method and then to apply the Arnold-Moser theorem, Birkhoff normal form of the Hamiltonian is computed followed by nonlinear stability of the equilibrium points is examined. Similar to the case of classical problem, we have found that in the presence of assumed perturbations, there always exists one value of mass parameter within the stability range at which the discriminant $D_4$ vanish, consequently, Arnold-Moser theorem fails, which infer that triangular equilibrium points are unstable in nonlinear sense within the stability range.  Present analysis is limited up to linear effect of the perturbations, which will be helpful to study the more generalized problem.
\end{abstract} 
 
\section{Introduction}
\label{sec:intro}

Since the time of Poincar\'{e}, invariant objects are very much important to understand the behavior of a dynamical system, especially, phase space. Moreover, there are many possible approaches to find the invariant objects, whereas the normal forms (truncated) are very useful because these can give integrable approximations to the dynamics under appropriate hypothesis \citep{jorba}. Because of the approximation of true dynamics by the normal forms, invariant objects of the initial system get approximated also, accordingly \citep{1995ASIB..336..343S,1998PhyD..114..197J}. The approximate first integrals are those quantities, which are almost preserved through the system's flow.  This shows that the surface levels by the flow are almost invariant. Some informations about the dynamics can be obtained through this property. To minimize the overflow and complexity in the computations, an appropriate approach is to use of power series or Fourier sires, or a combination of both to represent the object. Because in many cases they needed only a few numbers of terms to maintain the good accuracy. Some other approach can also be found in \cite{1998AcAau..43..493G,1999PhyD..132..189J}, in which trigonometric series is used. 
The normal forms of the Hamiltonian system up to some finite order is necessary to study the nonlinear stability of the equilibrium points using Arnold-Moser theorem in non-resonance case. They also help to know the behavior of dynamics in the neighborhood of the invariant objects. Many researchers have described the different method to find the normal forms of the Hamiltonian of the dynamical system \citep{poincarea,brikhoff,Deprit,ushikib,coppola1989ZaMM...69..275C,jorba,kishor2017Ap&SS.362..156K}. In the normal forms, the central idea is to find suitable transforms of the phase co-ordinates, which can convert the Hamiltonian system in its simplest form up to a finite order of accuracy. Normalization of Hamiltonian is obtained to change the Hamiltonian into its simplest form using the method of Lie transforms \citep{coppola1989ZaMM...69..275C,jorba}. 

Because of radiating primary in the present problem under the analysis, force due to radiation pressure came into existence \citep{Schuerman1980ApJ...238..337S,Ragos1993Ap&SS.209..267R}, which acts in opposite direction to the gravitational attraction force of the primary. Concept of Poynting-Roberston drag is came into the picture when, \cite{Poynting1903MNRAS..64A...1P} investigated the effect of radiation pressure on the moving particle in interplanetary space and  \cite{Robertson1937MNRAS..97..423R} modified the Poynting's theory through the principle
of relativity. In the analysis of Roberston, he considered only first order terms in the expression related to the ratio of velocity of the particle to that of the light. The radiation force is expressed as
\begin{eqnarray}
    \vec{F}&=&F_p\left(\frac{\vec{R}}{R}-\frac{\vec{V}.\vec{R}\vec{R}}{c R^2}-\frac{\vec{V}}{c}\right),\label{eq:prd}
  \end{eqnarray}
where $F_p$ is the radiation pressure force due to radiating primary; $\vec{R}$ is the position
vector of the particle relative to the radiating primary; $\vec{V}$ is the velocity
of the particle; and $c$ is the speed of the light. First term of the equation (\ref{eq:prd}) 
denotes the radiation pressure, second term represents the Doppler
shift due to the motion of the particle, whereas third term corresponds to the absorption
and subsequent re-emission part of induced radiation. The combined form of the last two terms
of the equation (\ref{eq:prd}) known as Poynting-Robertson (P-R) drag. \cite{Chernikov1970SvA....14..176C} analyzed the photogravitational restricted three body problem (RTBP) with P-R drag under the frame of 
Sun-planet-particle system and found that  non-collinear
(triangular) equilibrium points are unstable. Effect of P-R drag including radiation pressure is described by \cite{Schuerman1980ApJ...238..337S}. A similar analysis is presented by \cite{Murray1994Icar..112..465M} and \cite{Ragos1995A&A...300..568R} to observed the effect of P-R drag in the context of existence and stability of the equilibrium points. \cite{Kushvah2007Ap&SS.312..279K} examined the
nonlinear stability in the generalized photogravitational RTBP with P-R drag of first primary and oblateness of secondary and found that triangular equilibrium points are unstable, whereas \cite{vivek} investigated about the stability of non-collinear equilibrium points in the photogravitational elliptic RTBP with P-R drag. \cite{Kushvah2012Ap&SS.337..115K} and \cite{Kishor2013MNRAS.436.1741K} have analyzed the effect of radiation pressure force on the existence and linear stability of the equilibrium points in the generalized photogravitational Chermnykh-like problem with a disc. They found that the effect of perturbation factors are significant. In literature, many researchers have analyzed the photogravitational RTBP in nonlinear sense by considering one or two perturbations at a time \citep{McKenzie1981CeMec..23..223M,Ishwar1997CeMDA..65..253I,SubbaRao1997CeMDA..65..291S,Lhotka2015Icar..250..249L,Alvarez2015Ap&SS.358....3A} but very few of them have considered the problem under the combined influence of few perturbations \citep{Kushvah2007Ap&SS.312..279K,kishor2017Ap&SS.362..156K}. \cite{Ishwar2012Ap&SS.337..563I} have discussed about the nonlinear stability of out of plane equilibrium points in the RTBP with oblate primary and found that $L_6$ point is stable in nonlinear sense. \cite{xavier} have obtained diagonalized form of the Hamiltonian with P-R drag. \cite{kishor2017Ap&SS.362..156K} have studied nonlinear stability of triangular equilibrium points in the Chermnykh-like problem, in the presence of radiation pressure, oblateness and a disc. They found that these perturbations affect the numerical results significantly. 

Due to above reasons in addition to wide applications of the RTBP in mission design, we are motivated to study the problem under the influence of the radiation pressures and P-R drags of both primary and secondary. In the present study, we are interested to compute the fourth order normalized Hamiltonian and utilizing them to analyze the nonlinear stability of triangular equilibrium points using Arnold-Moser theorem in non-resonance case. Because of both primary and secondary radiating, the problem under analysis includes the four perturbing parameters in the form of mass reduction factors $q_1,\,q_2$ due to the radiation pressures of the primaries and P-R drags $W_1,\,W_2$ of both the primaries, respectively. 
The paper is organized as follows: In Section-\ref{sec:mf}, we have formulated the problem and found the equations of motion. Section-\ref{sec:sonf} presents the second order normalized Hamiltonian of the problem under analysis.  Nonlinear stability analysis is discussed in Section-\ref{sec:nlstb}. Section-\ref{sec:fonf} is devoted to Birkhoff normal form  and application of Arnold-Moser theorem in non-resonance case. Results are concluded in Section-\ref{sec:con}. For algebraic and numerical computations, Mathematica\textregistered \citep{wolfram2003mathematica} software package is used. The results of this study may be used to describe more generalized problem under the influence of other perturbations such as albedo, solar wind drag, Stokes drag etc. \citep{idrisi2018non,Singh2019Ap&SS.364....6S}.

\section{Mathematical Formulation}
\label{sec:mf}

We consider the photogravitational restricted three body problem with P-R drag, which consists of motion of an infinitesimal mass under the influence of gravitational field and radiation effect of two massive and radiating bodies of masses $m_1$ and $m_2,\, (m_1>m_2)$, respectively, called primaries. Forces, which govern the motion of infinitesimal mass are gravitational attractions,
radiation pressures and P-R drags of both the primaries, respectively.  It is assumed that gravitational effect of infinitesimal
mass on the system is negligible. Units are normalized such as units of mass and distance are taken as the sum of the masses of both the
primaries and separation distance between them, respectively, whereas unit of time is the time period of the rotating frame. We suppose that the coordinate of 
the primaries are $(-\mu,\,0),\, (1-\mu,\, 0),$ respectively and that of infinitesimal mass is $(x,\,y)$, then the equations of motion \citep{xavier} are 
\begin{eqnarray}
 \ddot{x}-2\dot{y}&=&\frac{\partial U}{\partial x},\label{x}\\
\ddot{y}-2\dot{x}&=&\frac{\partial U}{\partial y},\label{y}
\end{eqnarray}
where 
\begin{eqnarray}
\frac{\partial U}{\partial x}&=&x-\frac{q_1(1-\mu)(x+\mu)}{r_{1}^{3}}-\nonumber\\&&\frac{q_2(1-\mu)(x+\mu-1)}{r_{2}^{3}}
-\nonumber\\&&\frac{W_1S_1}{r_{1}^2}-\frac{W_2S_2}{r_{2}^2},\label{eq:Ux}\\
\frac{\partial U}{\partial y}&=&y-\frac{q_1(1-\mu)y}{r_{1}^{3}}-\frac{q_2(1-\mu)y}{r_{2}^{3}}\nonumber\\&&
-\frac{W_1S_3}{r_{1}^2}-\frac{W_2S_4}{r_{2}^2},\label{eq:Uy}
\end{eqnarray}
and further 
\begin{eqnarray*}
S_1&=&\frac{(x+\mu)\{(x+\mu)\dot{x}+y\dot{y}\}}{r_{1}^2}+\dot{x}-y,\\
S_2&=&\frac{(x+\mu-1)\{(x+\mu-1)\dot{x}+y\dot{y}\}}{r_{2}^2}+\dot{x}-y,\\
S_3&=&\frac{y\{(x+\mu)\dot{x}+y\dot{y}\}}{r_{1}^2}+\dot{y}+x+\mu,\\
S_4&=&\frac{y\{(x+\mu-1)\dot{x}+y\dot{y}\}}{r_{2}^2}+\dot{y}+x+\mu-1
\end{eqnarray*}
with $q_i=1-F_{pi}/F_{gi},\,i=1,2$ as mass reduction factors of both the primaries, respectively;  $F_{pi},\,F_{gi},\, i=1,2$ -are the forces of radiation pressure and gravitational attraction of the respective primaries; $W_1=[(1-q_1)(1-\mu)]/c_d$ and $W_2=[(1-q_2)\mu]/{c_d}$ as P-R drags of  both the primaries, respectively; $c_d$ -is the speed of light in non-dimensional form; $r_1,\,r_2$ - are distances of infinitesimal mass from the first and second primary, which are given as
\begin{eqnarray}
 r_{1}^2=(x+\mu)^2+y^2,\quad r_{2}^2=(x+\mu-1)^2+y^2.\label{eq:r1r2}\end{eqnarray}
The co-ordinates $(x_0,\,\pm y_0)$ of triangular equilibrium points $L_{4,5}$ are obtained on similar basis as in \cite{xavier}. To overcome the complexity in the analysis, co-ordinates $x_0$ and $y_0$ are linearized with respected to $W_1,\,W_2,\epsilon_1,\,\epsilon_2$, keeping in mind that the perturbing parameters lie in $(0,\,1)$ so, take $q_1=1-\epsilon_1,\,q_2=1-\epsilon_2$, where $\epsilon_1,\,\epsilon_2$ are very small. The linearized co-ordinates $x_0$ and $y_0$ are
\begin{eqnarray}
 x_0&=&\frac{1}{2}-\mu-\frac{4W_1(2-\mu)}{3\sqrt{3}}-\nonumber\\&&\frac{4W_2(1+2\mu)}{3\sqrt{3}}-\frac{\epsilon_1}{3}+\frac{\epsilon_2}{3},\label{eq:x4}\\
y_0&=&\pm\left[\frac{\sqrt{3}}{2}+\frac{4W_1(2-3\mu)}{9}+\right.\nonumber\\&&\left.\frac{4W_2(1-3\mu)}{9}-\frac{\epsilon_1}{3\sqrt{3}}-\frac{\epsilon_2}{3\sqrt{3}}\right].\label{eq:y4}
\end{eqnarray} The plus sign corresponds to $L_4$, whereas minus sign corresponds to $L_5$.
The Hamiltonian function of the problem is written as 
\begin{eqnarray}
 H&=&p_x\dot{x}+p_y\dot{y}-\frac{\dot{x}^2+\dot{y}^2}{2}-\frac{x^2+y^2}{2}-\nonumber\\&&x\dot{y}+\dot{x}y-\frac{(1-\mu)q_1}{r_1}-\frac{\mu q_2}{r_2}-\nonumber\\&&W_1S_5-W_2S_6,\label{eq:h}
\end{eqnarray}where
\begin{eqnarray*}
 S_5&=&\frac{(x+\mu)\dot{x}+y\dot{y}}{2r_{1}^2}-\arctan{\left(\frac{y}{x+\mu}\right)},\\
S_6&=&\frac{(x+\mu-1)\dot{x}+y\dot{y}}{2r_{1}^2}-\arctan{\left(\frac{y}{x+\mu-1}\right)}.
\end{eqnarray*}
The conjugate momenta $p_x,\,p_y$ corresponding to generalized co-ordinate $x,\,y$ respectively, are given as 
\begin{eqnarray}
 p_x&=&\dot{x}-y+\frac{W_1(x+\mu)}{2r_{1}^2}+\frac{W_2(x+\mu-1)}{2r_{2}^2},\label{eq:px}\\
p_y&=&\dot{y}+x+\frac{W_1 y}{2r_{1}^2}+\frac{W_2 y}{2r_{2}^2}.\label{eq:px}
\end{eqnarray}

\section{Second Order Normal Form of the Hamiltonian}
\label{sec:sonf}

 In the present analysis only the stability of $L_4$ is analyzed, because the dynamics of $L_5$ is similar to that of $L_4$. Only first order terms in the perturbing parameters $W_1,\,W_2,\,q_1,\,q_2$ are considered for simplifying the complex calculations involved in the problem through out the analysis. The second order normal form of the Hamiltonian of the problem under analysis is obtained in \cite{xavier} and for self sufficiency of this paper, we have taken some necessary expressions in appropriate form there to use under this section.
 Shifting the origin to the triangular equilibrium point $L_4$ using simple transformations as
\begin{eqnarray*}
&& x^*=x-x_0,\, y^*=y-y_0,\,\\&&{p_x}^*=p_x+y_0,\, {p_y}^*=p_y-x_0.\end{eqnarray*}
Substituting these variables in Hamiltonian (\ref{eq:h}), we get new Hamiltonian $H^*$. Now, expanding the new Hamiltonian using Taylor's series about the origin, which is now, the triangular equilibrium point, $H^*$ can be written as
\begin{eqnarray}
H^*=H_{0}^*+H_{1}^*+H_{2}^*+H_{3}^*+\dots+H_{n}^*+\dots,\label{eq:hs}\end{eqnarray}
where \begin{eqnarray}
H_{n}^*=\sum H_{ijkl}{x^*}^i{y^*}^j{{p_x}^*}^k{{p_y}^*}^l,\label{eq:hsn}\end{eqnarray}
such that $i+j+k+l=n$.
Since, the origin is the triangular equilibrium point, $H_{1}^*$ must vanish, whereas $H_{0}^*$ is constant hence,  it can be dropped out as it is irrelevant to the dynamics. The quadratic Hamiltonian $H_{2}^*$, which is to be normalized first and then to be used for higher order normalization, is given as
\begin{eqnarray}
H_{2}^*&=&\frac{{{p_x}*}^2+{{p_y}^*}^2}{2}+y^*{p_x}^*-x^*{p_y}^*\nonumber\\&&
+E{x^*}^2+Gx^*y^*+F{y^*}^2,\label{eq:h2}\end{eqnarray}
where
\begin{eqnarray}
E&=&\frac{1}{8}+\frac{4W_1}{\sqrt{3}}+\frac{2W_1}{\sqrt{3}}+\frac{\epsilon_1}{4}-\frac{\epsilon_2}{2},\label{eq:E}\\
F&=&-\frac{5}{8}-\frac{4W_1}{\sqrt{3}}-\frac{2W_1}{\sqrt{3}}-\frac{\epsilon_1}{4}+\frac{\epsilon_2}{2},\label{eq:F}\\
G&=&-\gamma\left(1-\frac{32W_1}{9\sqrt{3}}-\frac{16W_1}{9\sqrt{3}}\right.\nonumber\\&&\left.-\frac{2\epsilon_1}{9}+\frac{4\epsilon_2}{9}\right),\label{eq:G}\\
\text{with}\quad \gamma&=&\frac{3\sqrt{3}}{4}(1-2\mu).\label{eq:gama}
\end{eqnarray}
In the present study, the problem is dealt with four perturbation parameters in the form of P-R drag and radiation pressure of both the primaries. Hence, the coefficient $H_{ijkl}$ for $i,\,j,\,k,\,l=0,\,1,\,2,\,3,\,4$ such that $i+j+k+l=4$ in (\ref{eq:hsn}) can be bifurcated into five parts such as $H_{ijkl1},\,H_{ijkl2},\,H_{ijkl3},\,H_{ijkl4}$, and $H_{ijkl5}$, which corresponds to the terms in classical case, terms with P-R drag of first primary $W_1$, P-R
drag of second primary $W_2$, radiation pressure of first primary $\epsilon_1=1-q_1$ and radiation pressure of second primary $\epsilon_2=1-q_2$, respectively. Thus,
\begin{eqnarray} H_{ijkl}&=&H_{ijkl1}+H_{ijkl2}+H_{ijkl3}\nonumber\\&&+H_{ijkl4}+H_{ijkl5}.\label{eq:chsn}\end{eqnarray}
It is noted that if there is no perturbations in the system, i.e. $W_1=W_2=\epsilon_1=\epsilon_2=0$, then $H_{ijkl}=H_{ijkl1}$, which is nothing but the coefficient of the Hamiltonian in classical case.

Hamiltonian equations of motion of the infinitesimal mass in matrix form is written as
\begin{eqnarray}
\begin{bmatrix}\dot{x^*}\\\dot{y^*}\\\dot{{p_x}^*}\\\dot{{p_y}^*}\end{bmatrix}=\begin{bmatrix}0&1&1&0\\-1&0&0&1\\-2E&-G&0&1\\-G&-2F&-1&0\end{bmatrix}\begin{bmatrix}x^*\\y^*\\{p_x}^*\\{p_y}^*\end{bmatrix}.\label{eq:he}
\end{eqnarray}
The characteristic equation of the system (\ref{eq:he}) is
\begin{eqnarray} &&\lambda^4+2(E+F+1)\lambda^2+(4EF-G^2-\nonumber\\&&2E-2F+1)=0.\label{eq:ce}\end{eqnarray}
Solving the simplified discriminant of the characteristic equation (\ref{eq:ce}) as
\begin{eqnarray}(E+F+1)^2-(4EF-G^2-2E-2F+1)=0,\label{eq:dic}\end{eqnarray}
we have the value of critical mass ratio $0<\mu_c\leq (1/2)$ as 
\begin{eqnarray}\mu_c&=&0.0385209+0.0823761 W_1+0.0823761 W_2\nonumber\\&&+0.0178349 \epsilon_1-0.356699 \epsilon_2,\label{eq:cms}\end{eqnarray} which is similar to that of \cite{Kushvah2007Ap&SS.312..279K} and \cite{Kishor2013MNRAS.436.1741K} and agree with the classical value $\mu_c=0.0385209$. 
\begin{figure}
\centering
\includegraphics[width=\columnwidth]{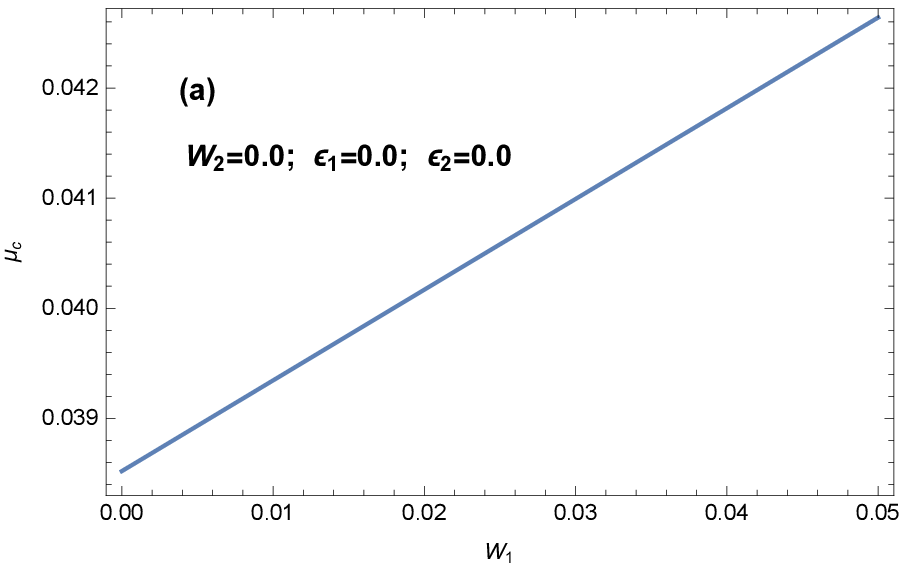}\\\includegraphics[width=\columnwidth]{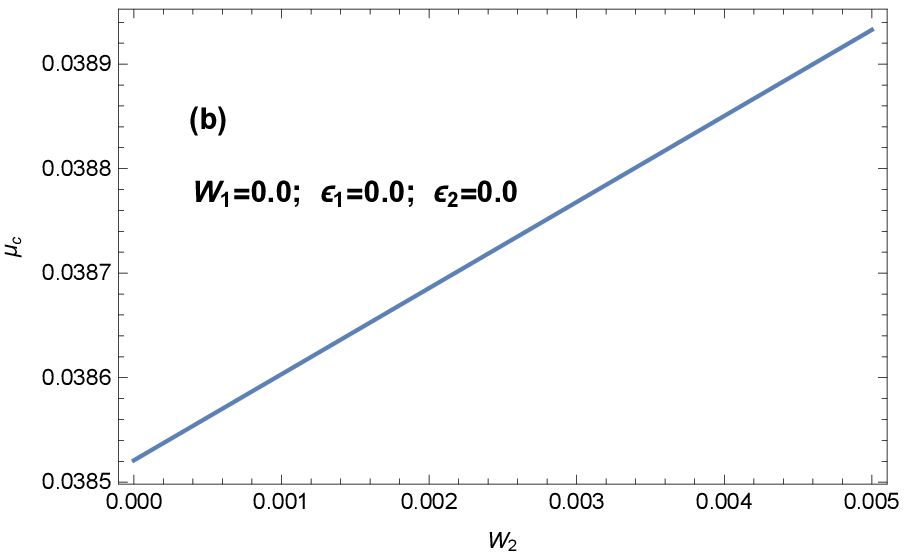}\\\includegraphics[width=\columnwidth]{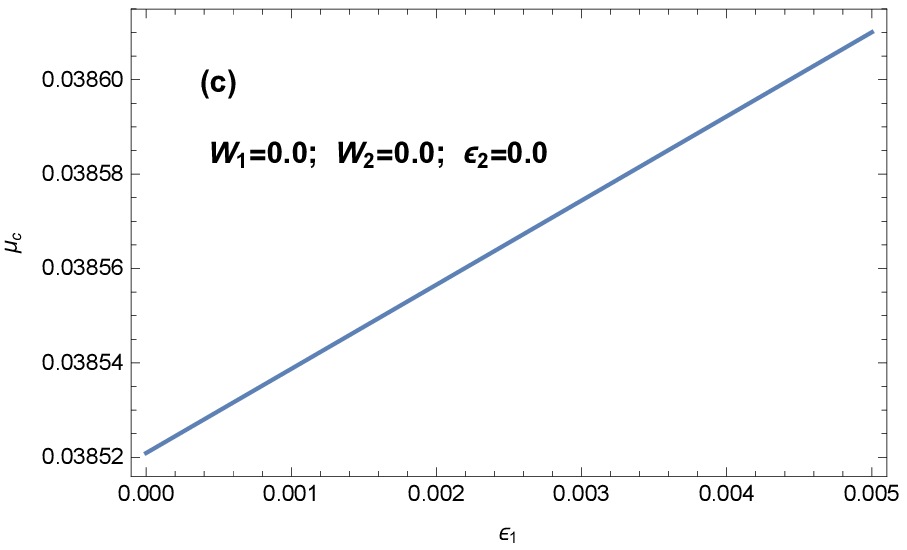}\\\includegraphics[width=\columnwidth]{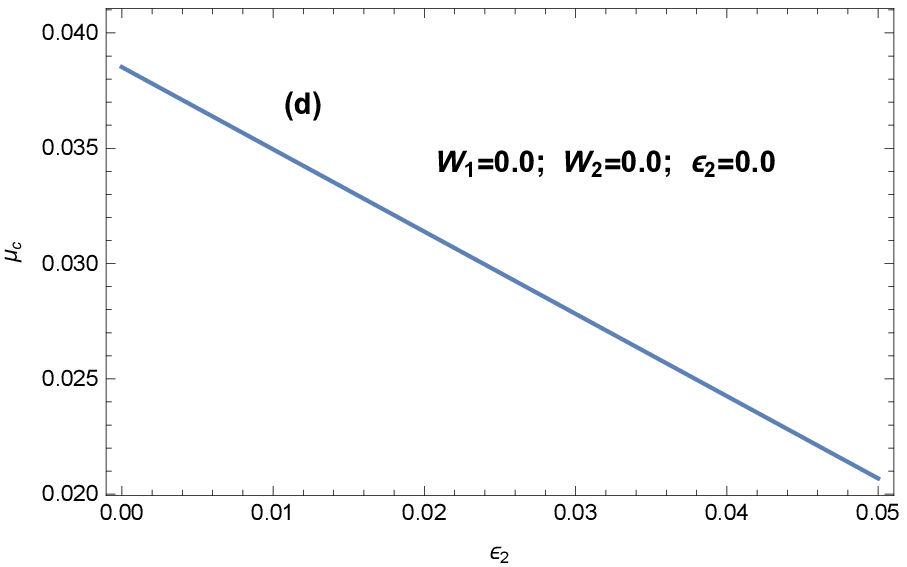}
   \caption{Variation of critical mass ratio $\mu_c$ with respect to (a) $W_1$, (b) $W_2$, (c) $\epsilon_1$ and (d) $\epsilon_2$.} \label{fig:muc}
\end{figure}
Figure (\ref{fig:muc})(a-d) shows the variations of critical mass ratio $\mu_c$ with respect to perturbing parameters $W_1,\,W_2,\, \epsilon_1$ and $\epsilon_2,$ respectively. We observed that the effects of the perturbations in question are significant.  As, system will be stable when four roots of the characteristic equation (\ref{eq:ce}) are pure imaginary, which is possible when  the mass parameter $\mu$ satisfy the condition $0<\mu<\mu_c$. Since, we are analyzing the nonlinear stability within the range of linear stability $0<\mu<\mu_c$, it is obvious to assume that roots of the characteristic equation (\ref{eq:ce}) are pure imaginary. Suppose, the roots of the characteristic equation (\ref{eq:ce}) are $\pm i\omega_1$ and $\pm i\omega_2$, where $\omega_1,\,\omega_2$ can be obtained by solving the equation
\begin{eqnarray} &&\omega^4-2(E+F+1)\omega^2+(4EF-G^2-\nonumber\\&&2E-2F+1)=0.\label{eq:cew}\end{eqnarray}
Motion corresponds to frequencies $\omega_{1},\,\omega_2\in \mathbb{R}$ are known as long and short periodic motion of infinitesimal mass at $L_4$ with periods of $2\pi/\omega_1$ and $2\pi/\omega_2$, respectively. Frequencies $\omega_{1},\,\omega_2$ corresponding to the long and short periodic motion are related to each other by the means of relations
\begin{eqnarray} \omega_{1}^2+\omega_{2}^2&=&2E+2F+2,\label{eq:fqr1}\\\omega_{1}^2\omega_{2}^2&=&4EF-G^2-2E-2F+1.\label{eq:fqr2}\end{eqnarray}
Substituting the values of $E,\,F$ and $G$ from equations (\ref{eq:E}-\ref{eq:G}), we get
\begin{eqnarray} \omega_{1}^2+\omega_{2}^2&=&1,\label{eq:fqr3}\\\omega_{1}^2\omega_{2}^2&=&\frac{27}{16}\gamma^2-4\sqrt{3}W_1-2\sqrt{3}W_2-\nonumber\\&&\frac{2\epsilon_1}{3}+\frac{3\epsilon_2}{4},\label{eq:fqr4}\end{eqnarray}
where the values of $\omega_1$ and $\omega_2$ are
\begin{eqnarray}
 \omega_1=\sqrt{-1+\sqrt{1-4\delta}},\quad \omega_2=\sqrt{-1-\sqrt{1-4\delta}},\label{eq:frq}
\end{eqnarray}with
\begin{eqnarray}
 \delta=\frac{27}{16}-\gamma^2-4\sqrt{3}W_1-2\sqrt{3}W_2-\frac{3\epsilon_1}{4}+\frac{3\epsilon_2}{2}.\label{eq:delta}
\end{eqnarray}

The real normalized Hamiltonian of the Hamiltonian (\ref{eq:h2}) up to second order is given as \citep{xavier}
\begin{eqnarray}
 H_2=\omega_1\frac{\bf{x^2+p_{x}^2}}{2}+\omega_2\frac{\bf{y^2+p_{y}^2}}{2},\label{eq:rnh}
\end{eqnarray} which is complexified by using the co-ordinate transformations
\begin{eqnarray}
\bf{x}&=&\frac{X+iP_X}{\sqrt{2}},\\
\bf{y}&=&\frac{-Y+iP_Y}{\sqrt{2}},\\
\bf{p_x}&=&\frac{iX+P_X}{\sqrt{2}},\\
\bf{p_y}&=&\frac{iY-P_Y}{\sqrt{2}}
\end{eqnarray} and changed as
\begin{eqnarray}H_2=i\omega_1 XP_X-i\omega_2YP_Y,\label{eq:nh2}\end{eqnarray}
Finally, symplectic matrix $\bf{C}$ of the  symplectic transformations, which are used to obtain the complex normal form of Hamiltonian is given as \citep{xavier}
\begin{eqnarray}&&{\bf{C}}=\begin{bmatrix}s_{ij}\end{bmatrix},\,1\leq i,j\leq 4\label{eq:sm}\end{eqnarray}
with
\begin{eqnarray*}&&s_{11}=0=s_{12},\,s_{13}=\frac{1-2F+\omega_{1}^2}{\sqrt{d(\omega_1)}},\\&& s_{14}=\frac{1-2F+\omega_{1}^2}{\sqrt{d(\omega_2)}},\,s_{21}=\frac{2\omega_1}{\sqrt{d(\omega_1)}},\\&& s_{22}=\frac{2\omega_1}{\sqrt{d(\omega_2)}},\, s_{23}=\frac{G}{\sqrt{d(\omega_1)}},\\&&s_{24}=\frac{G}{\sqrt{d(\omega_2)}},\, s_{31}=\frac{\omega_{1}^3-(2F+1)\omega_1}{\sqrt{d(\omega_1)}},\\&& s_{32}=\frac{\omega_{2}^3-(2F+1)\omega_2}{\sqrt{d(\omega_2)}},\, s_{33}=\frac{-G}{\sqrt{d(\omega_1)}},\\&& s_{34}= \frac{-G}{\sqrt{d(\omega_2)}},\, s_{41}=\frac{G\omega_1}{\sqrt{d(\omega_1)}},\, s_{42}=\frac{G\omega_2}{\sqrt{d(\omega_2)}},\\&& s_{43}=\frac{1-2F-\omega_{1}^2}{\sqrt{d(\omega_1)}},\, s_{44}=\frac{1-2F-\omega_{2}^2}{\sqrt{d(\omega_2)}},\end{eqnarray*}
where $d(\omega_i)$ for $i=1,\,2$ is obtained from the following equation
\begin{eqnarray} d(\omega)&=&\omega\left[\omega^4-(2E+6F)\omega^2+\right.\nonumber\\&&\left.(4EF+4F^2-2E+2F-2)\right].\label{eq:dw}\end{eqnarray}
\section{Nonlinear Stability in Non-resonance Case}
\label{sec:nlstb}
Nonlinear stability of the equilibrium points can be described in two cases, one as resonance case and other as non-resonance case. For resonance case, the nonlinear stability is studied through the theorems of \cite{Markeev1977SvA....21..507M} as in \cite{K.Gozdzieski1998CeMDA..70...41G} and for non-resonance case, it is analyzed through the Arnold-Moser theorem. In the present analysis the nonlinear stability of the perturbed triangular equilibrium point in non-resonance case will be studied through Arnold-Moser theorem \citep{Mayer,Hall1992ihds.book.....M}, which is  described as follows:

Consider the Hamiltonian expressed in action variables $ I_1,\, I_2$ and angles  variables $\phi_1,\, \phi_2$  as,
\begin{eqnarray} &&K=K_2+ K_4 + \dots +K_{2m}+K_{2m+1}, \label{eq:amh}\end{eqnarray}
in which: (i) $K_{2m}$ is homogeneous polynomial of degree $m$ in action variables $I_1,\,I_2$ and $K_{2m+1}$ is higher degree polynomial than $m$ (ii)  $K_2=\omega_1 I_1-\omega_2 I_2$ with $\omega_{1,2}$ as positive constants (iii)  $K_4=-(AI_{1}^2+BI_1I_2+CI_{2}^2)$, where  $A,\,B,\,C$  are constants to be determined. Since, $K_2,\,K_4,\,\dots,K_{2m}$ are functions of $I_1$ and $I_2$, the Hamiltonian (\ref{eq:amh}) follows the  Birkhoff normal form \citep{brikhoff} up to the terms $m$. This can be obtained with some non-resonance condition on the frequencies $\omega_{1},\,\omega_2$. To state the Arnold-Moser theorem, we assume that $K$ is in the required form.

{\bf{Arnold-Moser Theorem:}} \textit{The origin is stable for the system whose Hamiltonian is (\ref{eq:amh}) provided    for    some  $\nu,\,\,2\leq\nu\leq m$,    $D_{2\nu} = K_{2\nu}(\omega_2,\omega_1) \neq 0$}.

Since, for Arnold-Moser theorem, Birkhoff normal form of the Hamiltonian is necessary and for Birkhoff normal form, assumption of non-resonance on frequencies is required. The non-resonance condition of frequencies as in \cite{Deprit1967AJ.....72..173D,kishor2017Ap&SS.362..156K} is that if $\omega_{1},\,\omega_2$ are frequencies of infinitesimal mass in linear dynamics and $\sigma\in\mathbb{Z}$  such  that $\sigma\geq2$, then 
\begin{eqnarray}\sigma_1\omega_1+\sigma_2\omega_2\neq 0\label{eq:ic} \end{eqnarray}
for all $\sigma_{1},\,\sigma_2\in\mathbb{Z}$ satisfying $|\sigma_1|+|\sigma_2|\leq 2\sigma$. This is also, called as condition of irrationality, which insures that there exists a symplectic normalizing transformation which transform the Hamiltonian (\ref{eq:hs}) in  the form of Hamiltonian (\ref{eq:amh}). Coefficients of the normalized Hamiltonian are independent on the integer $\sigma$ as well as to the transformation obtained. In specific 
\begin{eqnarray}\text{det}\begin{vmatrix}\frac{\partial^2K}{\partial I_{1}^2}&\frac{\partial^2K}{\partial I_1\partial I_{2}}&\frac{\partial K}{\partial I_{1}}\\\frac{\partial^2K}{\partial I_2\partial I_{1}}&\frac{\partial^2K}{\partial I_{2}^2}&\frac{\partial K}{\partial I_{2}}\\\frac{\partial K}{\partial I_{1}}&\frac{\partial K}{\partial I_{2}}&0\end{vmatrix}_{I_1,I_2=0}\end{eqnarray} is invariant of the Hamiltonian (\ref{eq:amh}) with respect to the symplectic transformation considered. The nonlinear stability of perturbed triangular equilibrium points is analyzed through the Arnold-Moser theorem under these conditions. In classical case frequencies $\omega_{1},\,\omega_2$ satisfy the condition $0<\omega_2<(1/\sqrt{2})<\omega_1<1$. Therefore, if $\sigma=2$, then irrationality  condition (\ref{eq:ic}) fails for following pairs of integers $\sigma_1=1,\, \sigma_2=-2$, $\sigma_1=-1,\, \sigma_2=2$, $\sigma_1=1,\, \sigma_2=-3$  and $\sigma_1=-1,\, \sigma_2=3$. First, two pairs of integers with condition (\ref{eq:ic})   yield $(\omega_1/\omega_2)=(1/2)$ and last two pairs of integers give $(\omega_1/\omega_2)=(1/3)$, which are also known as second and third order resonance of the frequencies respectively.  If $(\omega_1/\omega_2)=(1/2)$ or $\omega_1=2\omega_2$, then from equations (\ref{eq:fqr3}-\ref{eq:fqr4}), we get
\begin{eqnarray} \frac{4}{25}=\frac{27}{16}\gamma^2-4\sqrt{3}W_1-2\sqrt{3}W_2-\frac{2\epsilon_1}{3}+\frac{3\epsilon_2}{4}.\label{eq:mc11}\end{eqnarray}
Simplifying equation (\ref{eq:mc11}), we have a quadratic equation in $\mu$ as
\begin{eqnarray}&&\frac{27}{16}\mu^2-\frac{27}{16}\mu+\left(\sqrt{3}W_1+\frac{\sqrt{3}W_2}{2}+\right.\nonumber\\&&\left.\frac{3\epsilon_1}{16}-\frac{3\epsilon_2}{8}+\frac{1}{25}\right)=0.\label{eq:mc12}\end{eqnarray}
The solution $\mu=\mu_{c1}$ of equation (\ref{eq:mc12}) within the stability range $0<\mu<\mu_c$ is
\begin{eqnarray}\mu_{c1}&=&0.0242939+1.078820 W_1+0.539409 W_2\nonumber\\&&+0.116785 \epsilon_1-0.233571 \epsilon_2.\label{eq:mc13}\end{eqnarray}
This means, Arnold-Moser theorem fails at  $\mu_{c1}\in (0,\,\mu_c)$. If $(\omega_1/\omega_2)=(1/3)$ or $\omega_1=3\omega_2$, then proceeding on similar basis, we find that Arnold-Moser theorem fails at $\mu=\mu_{c2}$, where
\begin{eqnarray}\mu_{c2}&=&0.013516+1.054920 W_1+0.527459 W_2\nonumber\\&&+0.114198 \epsilon_1-0.228396 \epsilon_2.\label{eq:mc23}\end{eqnarray}
Equations (\ref{eq:mc13}-{\ref{eq:mc23}) are similar to that of the results in \cite{Deprit1967AJ.....72..173D,Kishor2013MNRAS.436.1741K} and agree with classical result in the absence of perturbing parameters.  To see the effects of perturbing parameters on $\mu_{c1}$ and $\mu_{c2}$, its numerical values are computed and presented in Table-\ref{tab:muc12}. From Table-\ref{tab:muc12}, it is clear that the values of $\mu_{c1}$ and $\mu_{c2}$ are very much affected from radiation pressures and P-R drags of the primaries.
	
	\begin{table}
\caption{ $\mu_{c1}$ and $\mu_{c2}$ at different values of perturbing parameters.}
\label{tab:muc12}
\begin{tabular}{@{}|rrrrrr|@{}} 
\hline
$W_1$& $W_1$& $\epsilon_1$&$\epsilon_2$ &$\mu_{c1}$&$\mu_{c2}$\\	
\hline
 0.000&0.0000&0.000&0.00&0.024294&0.013516\\
 &&&&&\\
 0.005&0.0000&0.000&0.00&0.029688&0.018791\\
 0.010&0.0000&0.000&0.00&0.035082&0.024065\\
 0.015&0.0000&0.000&0.00&0.040476&0.029340\\
 &&&&&\\
 0.000&0.0040&0.000&0.00&0.026452&0.015626\\
 0.000&0.0045&0.000&0.00&0.026721&0.015890\\
 0.000&0.0050&0.000&0.00&0.026991&0.016153\\
 &&&&&\\
 0.000&0.0000&0.001&0.00&0.024411&0.013630\\
 0.000&0.0000&0.002&0.00&0.024528&0.013744\\
 0.000&0.0000&0.003&0.00&0.024644&0.013859\\
 &&&&&\\
 0.000&0.0000&0.000&0.01&0.021958&0.011232\\
 0.000&0.0000&0.000&0.02&0.019623&0.008948\\
 0.000&0.0000&0.000&0.03&0.017287&0.006664\\
 &&&&&\\
 0.005&0.0040&0.001&0.01&0.029627&0.018731\\
\hline
\end{tabular}
\end{table}

\section{Fourth order Normalized Hamiltonian}
\label{sec:fonf}

Since, Birkhoff's normal form up to fourth order of the Hamiltonian is necessary to apply the Arnold-Moser theorem, which is computed from second order normalized Hamiltonian (\ref{eq:hsn}) using Lie transform method described in \cite{1988CeMec..45..103C,1989CeMec..45..103C,jorba,celletti2010stability,kishor2017Ap&SS.362..156K}. As, in the paper of  \cite{1989CeMec..45..103C} as well as in the book of \cite{celletti2010stability}, higher order normalized Hamiltonian is 
\begin{eqnarray}K&=&K_2+K_3+K_4+\dots+K_n+\dots,\label{eq:honh}\end{eqnarray} where 
\begin{eqnarray}K_n&=&\sum{K_{ijkl}X^iY^j{P_X}^k{P_Y}^l}\label{eq:kn}\end{eqnarray}
such that $i+j+k+l=n.$ 
Quadratic part of $K$ is $K_2=H_2$, whereas $K_n$ through the nth step of Lie transform is given as 
\begin{eqnarray}K_n&=&\frac{1}{n}\{H_2,\,G_n\}+(\text{known terms}),\label{eq:gn}\end{eqnarray} where Lie bracket of normalized quadratic Hamiltonian $H_2$ and generating function $G_n$ is defined as
\begin{eqnarray}\{H_2,\,G_n\}&=&\frac{\partial H_2}{\partial X}\frac{\partial G_n}{\partial P_X}-\frac{\partial H_2}{\partial P_X}\frac{\partial G_n}{\partial X}\nonumber\\&&
+\frac{\partial H_2}{\partial Y}\frac{\partial G_n}{\partial P_Y}-\frac{\partial H_2}{\partial P_Y}\frac{\partial G_n}{\partial Y}.\label{eq:lb}\end{eqnarray}
Using $H_2$ from equation (\ref{eq:hsn}), it reduces to
\begin{eqnarray}\{H_2,\,G_n\}&=&\mathbf{i}\omega_1\left(P_X\frac{\partial G_n}{\partial P_X}-X\frac{\partial G_n}{\partial X}\right)\nonumber\\&&
+\mathbf{i}\omega_2\left(P_Y\frac{\partial G_n}{\partial P_Y}-Y\frac{\partial G_n}{\partial Y}\right).\label{eq:lb1}\end{eqnarray}
The choice of generating function $G_n$ is such that the above partial differential operator on $G_n$ remove large possible number of terms from the expression of $K_n$. As, each terms of the $K_n$ is of the form $\alpha X^iY^j{P_X}^k{P_Y}^l$, where $\alpha$ is constant, we can assume terms in $G_n$ of the form $\beta X^iY^j{P_X}^k{P_Y}^l$, where constant $\beta$ is to be determined. Therefore,  we obtain that
\begin{eqnarray}\frac{\{H_2,\,G_n\}}{n}=\frac{\mathbf{i}\beta}{n}\left[(k-i)\omega_1-(l-j)\right]X^iY^j{P_X}^k{P_Y}^l,\label{eq:gn0}\end{eqnarray}
and hence, \begin{eqnarray}\beta=\frac{\mathbf{i}\alpha}{\left[(k-i)\omega_1-(l-j)\right]},\,\,i+j+k+l=n.\label{eq:bta}\end{eqnarray}
This shows that even in the non-resonance case, the term of the form $X^iY^j{P_X}^i{P_Y}^j$ in $K_n$ can not be deleted because of vanishing denominator in (\ref{eq:bta}) at $i=k,\,j=l$, whereas in the resonance case some additional non-removable terms occur while solving the generating function $G_n$. Hence, in non-resonance case, the Hamiltonian of the present problem can be written in the form of (\ref{eq:honh}), in which
\begin{eqnarray}K_2&=&i\omega_1XP_X -i\omega_2YP_Y,\label{eq:sonh}\\K_3&=&0,\label{eq:tonh}\\
K_4&=&\frac{AX^2P_{X}^2+BXP_XYP_Y+CY^2P_{Y}^2}{2},\label{eq:fonh}\end{eqnarray}
where $A=2K_{2020},\,B=2K_{1111},$ and $C=2K_{0202}$.
Using action variables $I_1=iXP_X$ and $I_2=iYP_Y$ in equations (\ref{eq:sonh}-\ref{eq:fonh}), we get 
\begin{eqnarray}K_2&=&\omega_1I_1 -\omega_2I_2,\label{eq:sonhav}\\K_3&=&0,\label{eq:tonhav},\\
K_4&=&-\left(AI_{1}^2+BI_I1_2+CI_{2}^2\right).\label{eq:fonhav}\end{eqnarray}
Thus, normalized Hamiltonian up to fourth order is 
\begin{eqnarray}K(I_1,\,I_2)&=&K_2+K_3+K_4\nonumber\\
&=&\omega_1I_1 -\omega_2I_2-K_{2020}I_{1}^2+\nonumber\\&&K_{1111}I_I1_2+K_{0202}I_{2}^2,\label{eq:bnf}\end{eqnarray} which agree with that of \cite{Deprit1967AJ.....72..173D,Kushvah2007Ap&SS.312..279K,kishor2017Ap&SS.362..156K}.

\begin{figure}
\centering
 \includegraphics[width=\columnwidth]{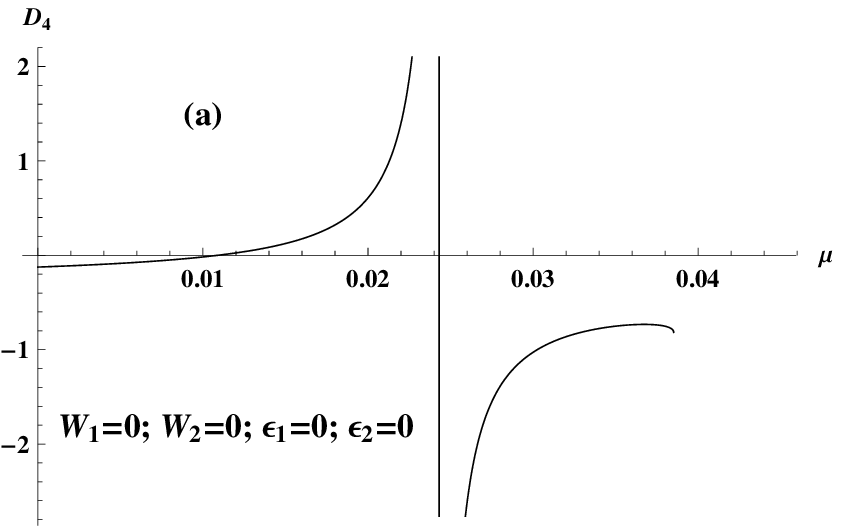}\\\includegraphics[width=\columnwidth]{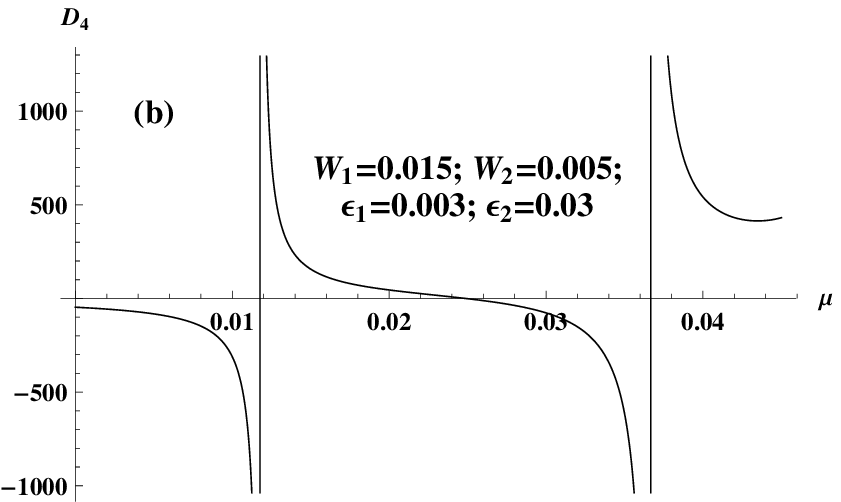}
   \caption{Zero $(\mu_0)$ of the determinant $D_4$ within the stability range $0<\mu<\mu_c$ at: (a) $W_1=W_2=\epsilon_1=\epsilon_2=0$ (classical case); (b) $W_1=0.015,\,W_2=0.005,\,\epsilon_1=0.003,\,\epsilon_2=0.03$ (perturbed case).\label{fig:d41}}
\end{figure}

\begin{figure}
\centering
 \includegraphics[width=\columnwidth]{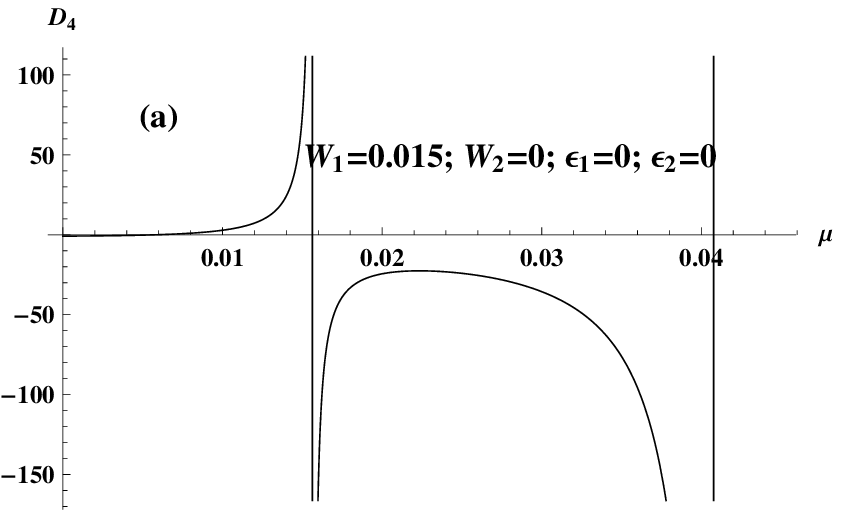}\\\includegraphics[width=\columnwidth]{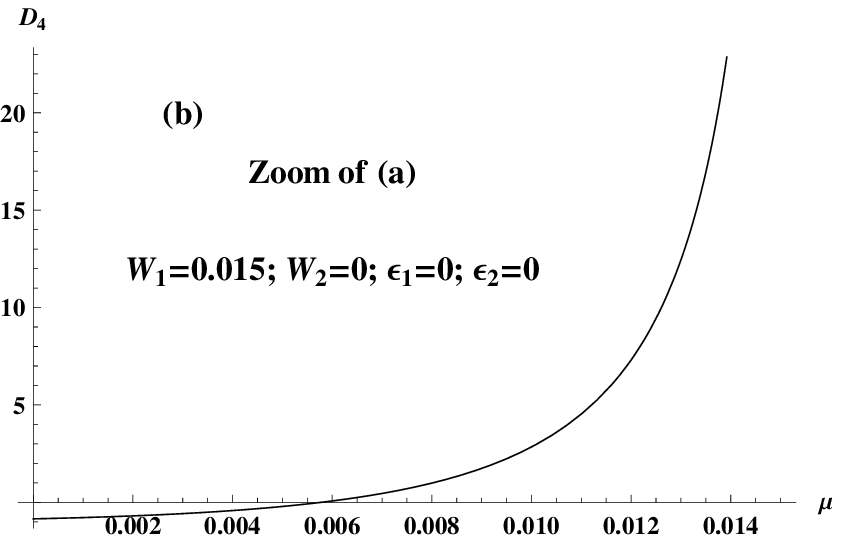}
   \caption{Zero $(\mu_0)$ of the determinant $D_4$ within the stability range $0<\mu<\mu_c$ at: (a) $W_1=0.015,\,W_2=\epsilon_1=\epsilon_2=0$ (only in presence of P-R drag of first primary); (b) Zoom of specified region of figure (a).}
  \label{fig:d42}
\end{figure}

\begin{figure}
\centering
 \includegraphics[width=\columnwidth]{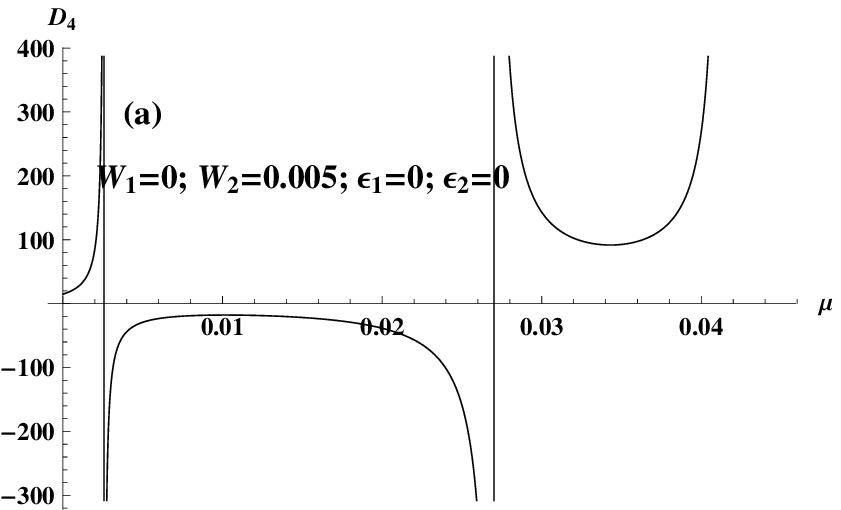}\\\includegraphics[width=\columnwidth]{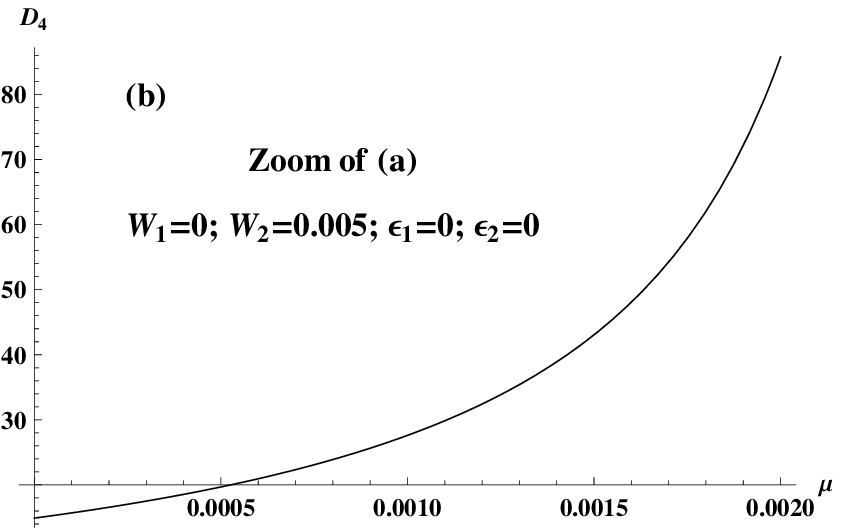}
   \caption{Zero $(\mu_0)$ of the determinant $D_4$ within the stability range $0<\mu<\mu_c$ at: (a) $W_2=0.005,\,W_1=\epsilon_1=\epsilon_2=0$ (only in presence of P-R drag of second primary); (b) Zoom of specified region of figure (a).}
  \label{fig:d43}
\end{figure}

\begin{figure}
\centering
 \includegraphics[width=\columnwidth]{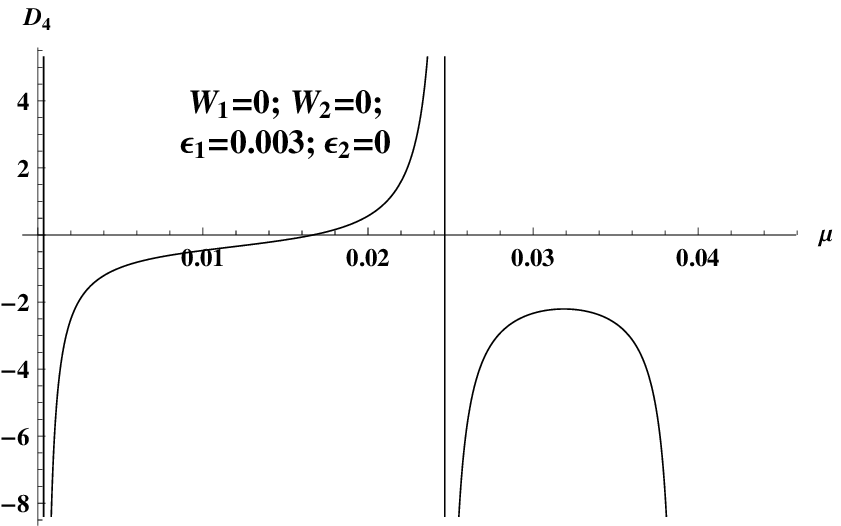}
   \caption{Zero $(\mu_0)$ of the determinant $D_4$ within the stability range $0<\mu<\mu_c$ at: $\epsilon_1=1-q_1=0.003,\,W_1=W_2=\epsilon_2=0$ (only in presence of radiation pressure of first primary).}
  \label{fig:d44}
\end{figure}

\begin{figure}
\centering
\includegraphics[width=\columnwidth]{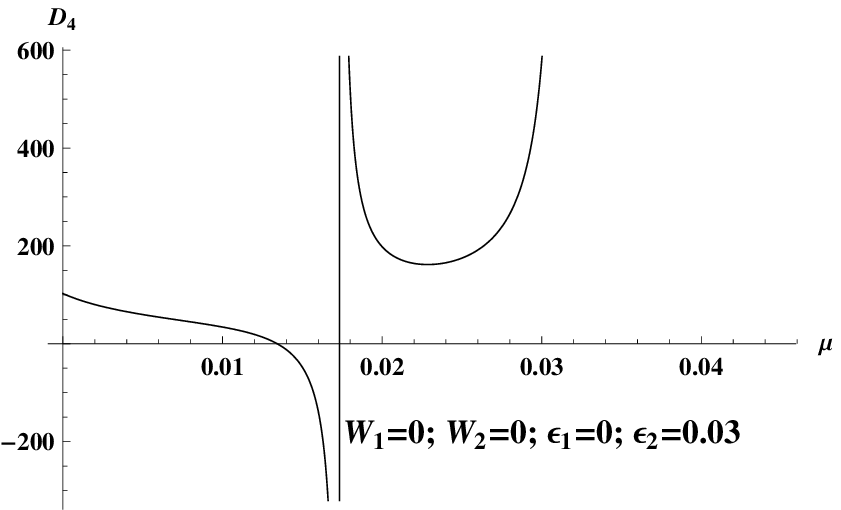}
   \caption{Zero $(\mu_0)$ of the determinant $D_4$ within the stability range $0<\mu<\mu_c$ at $\epsilon_2=1-q_2=0.03,\,W_1=W_2=\epsilon_1=0$ (only in presence of radiation pressure of second primary).}
  \label{fig:d45}
\end{figure}

 Form equations (\ref{eq:bnf}), it is clear that fourth order normalized Hamiltonian is the function of only action variables $I_1,\, I_2$, which shows that these are in Birkhoff normal form. The coefficients $K_{ijkl}$ used in the equation (\ref{eq:fonh} or \ref{eq:fonhav}) can be written into $5$ parts such as $K_{ijkl1}$, $K_{ijkl2}$, $K_{ijkl3}$, $K_{ijkl4}$ and $K_{ijkl5}$ for $i,\,j,\,k,\,l=0,\,1,\,2,\,3,\,4$ such that $i+j+k+l=4$. These coefficients corresponds to the term of classical part, terms with P-R drags $W_1$ and
 $W_2$ of first and second primary, radiation pressures  $\epsilon_1=1-q_1$ and $\epsilon_2=1-q_2$ of first and second primary, respectively. In the absence of perturbing parameters i.e. for $W_1=W_1=\epsilon_1=\epsilon_2=0$, $K_{ijkl}={K{ijkl1}}$. Therefore, $K_{2020},\,K_{1111}$ and $K_{0202}$ become   
\begin{eqnarray}K_{2020}&=&K_{20201}+K_{20202}+K_{20203}+\nonumber\\&&K_{20204}+K_{20205},\label{eq:k2020}\\
K_{1111}&=&K_{11111}+K_{11112}+K_{11113}+\nonumber\\&&K_{11114}+K_{11115},\label{eq:k1111}\\
K_{0202}&=&K_{02021}+K_{02022}+K_{02023}+\nonumber\\&&K_{02024}+K_{02025}\label{eq:k0202}.\end{eqnarray}
The algebraic expressions of above $15$ coefficients on right hand sides of equations (\ref{eq:k2020}-\ref{eq:k0202}) are too complicated and huge to be placed here hence, we avoid to present in the paper. These are utilized to compute  the determinant $D_{4}=K_4(\omega_2,\,\omega_1)$ for applying the Arnold-Moser theorem.  For the simplicity,  $D_4$ is expressed as 
\begin{eqnarray} D_4&=&\left(\frac{A_1} {B_1}\right) +\left(\frac{A_2} {B_2}\right) W_1+\left(\frac{A_3} {B_3}\right) W_2+\nonumber\\&&\left(\frac{A_4} {B_4}\right)  \epsilon_1+\left(\frac{A_5} {B_5}\right)  \epsilon_2,\label{eq:d4}\end{eqnarray}
where $A_i,\, B_i,\, i=1,2,3,4,5$  are numerator and denominator of the coefficients, which correspond to classical part, P-R drags $W_1$ and $W_2$ of the primaries,  radiation pressure $\epsilon_1=1-q_1$ and $\epsilon_2=1-q_2$ of the primaries, respectively.  On simplification, we found that  
\begin{eqnarray}A_1&=&-35+541\omega_{1}^2\omega_{2}^2-644\omega_{1}^4\omega_{2}^4,\quad\quad\quad\quad\label{eq:a1}\end{eqnarray}
\begin{eqnarray}A_2&=&26244\left(2262-653b\right)-\nonumber\\&&27\left(5292162-4787719 b\right)\omega_{1}^2\omega_{2}^2\nonumber\\&&-2\left(402982614-10430203 b\right)\omega_{1}^4\omega_{2}^4\nonumber\\&&+32\left(12457908-1490819b\right)\omega_{1}^6\omega_{2}^6\nonumber\\&&+1024\left(67581+1634b\right)\omega_{1}^8\omega_{2}^8,\label{eq:a2}\end{eqnarray}
\begin{eqnarray}A_3&=&-78732\left(416-241\sqrt{3} b\right)-\nonumber\\&&27\left(3181248+4414649sqrt{3} b\right)\omega_{1}^2\omega_{2}^2\nonumber\\&&-6\left(72610776+10390609\sqrt{3} b\right)\omega_{1}^4\omega_{2}^4\nonumber\\&&+32\left(6808752-212191\sqrt{3} b\right)\omega_{1}^6\omega_{2}^6\nonumber\\&&+1024\left(36828+997\sqrt{3} b\right)\omega_{1}^8\omega_{2}^8,\label{eq:a3}\end{eqnarray}
\begin{eqnarray}A_4&=&8748\left(195\sqrt{3}-584 b\right)-\nonumber\\&&27\left(465795\sqrt{3}-1556744 b\right)\omega_{1}^2\omega_{2}^2\nonumber\\&&+2\left(10722915\sqrt{3}-11609036 b\right)\omega_{1}^4\omega_{2}^4\nonumber\\&&+32\left(200970\sqrt{3}-103079 b\right)\omega_{1}^6\omega_{2}^6\nonumber\\&&-512\left(2565\sqrt{3}-3217 b\right)\omega_{1}^8\omega_{2}^8,\label{eq:a4}\end{eqnarray}
\begin{eqnarray}A_5&=&-8748\left(507\sqrt{3}-688 b\right)+\nonumber\\&&27\left(315819\sqrt{3}-1533728 b\right)\omega_{1}^2\omega_{2}^2\nonumber\\&&+2\left(3085838\sqrt{3}+759212 b\right)\omega_{1}^4\omega_{2}^4\nonumber\\&&-32\left(941526\sqrt{3}-283835 b\right)\omega_{1}^6\omega_{2}^6\nonumber\\&&-512\left(10251\sqrt{3}+877 b\right)\omega_{1}^8\omega_{2}^8,\label{eq:a5}\end{eqnarray}
\begin{eqnarray}B_1&=&8\left(1-4\omega_{1}^2\omega_{2}^2\right)\left(4-25\omega_{1}^2\omega_{2}^2\right),\label{eq:b1}\\
B_2&=&864ab,\label{eq:b2}\\
B_3&=&B_2,\label{eq:b3}\\
B_4&=&1152ab,\label{eq:b4}\\
B_5&=&576ab,\label{eq:b5}\end{eqnarray}
\begin{eqnarray}a&=&\left[\omega_{1}^2\omega_{2}^2\left(1-4\omega_{1}^2\omega_{2}^2\right)\right.\\&&\left.\left(4-25\omega_{1}^2\omega_{2}^2\right)\left(117+16\omega_{1}^2\omega_{2}^2\right)\right],\label{eq:a}\\
b&=&\sqrt{\left(27-16\omega_{1}^2\omega_{2}^2\right)}.\label{eq:b}\end{eqnarray}
In the absence of perturbing parameters, 
\begin{eqnarray}D_4&=&\frac{-35+541\omega_{1}^2\omega_{2}^2-644\omega_{1}^4\omega_{2}^4}{8\left(1-4\omega_{1}^2\omega_{2}^2\right)\left(4-25\omega_{1}^2\omega_{2}^2\right)}, \label{eq:cd}\end{eqnarray} which agree with the classical result \citep{Deprit1967AJ.....72..173D,Mayer,Kushvah2007Ap&SS.312..279K,kishor2017Ap&SS.362..156K}.
In order to analyze the nonlinear stability of triangular equilibrium points in non-resonance case using Arnold-Moser theorem, we plot the determinant $D_4$ with respect to the mass parameter $\mu$ to insure the value of $D_4=K_4(\omega_2,\omega_1)$. From figures (\ref{fig:d41}-\ref{fig:d45}), it is clear that within the linear stability range $0<\mu<\mu_c$, there exists one value of mass parameter $\mu=\mu_0$, called the zero of $D_4$, at which $D_4$ vanish in each case. Thus, Arnold-Moser theorem fails, which insure that in non-resonance case, triangular equilibrium points of the problem under analysis are unstable in nonlinear sense within the linear stability range $0<\mu<\mu_c$. To see the effect of perturbing parameters, we have computed values of the zero $(\mu_0)$ of $D_4$ and critical mass ratio $(\mu_c)$ at different values of perturbing parameters $W_1,\,W_2,\,\epsilon_1,\,\epsilon_2$ and results are placed in Table-\ref{tab:table1}. From Table-\ref{tab:table1}, it is noticed that on increase in the values of $W_1,\,W_1,\,\epsilon_1$, value of critical mass $\mu_c$ increases but the value of $\mu_0$ is nonzero in each case. On the other hand, on increase in the value of $\epsilon_2$, $\mu_c$ decreases with nonzero $\mu_0$. Thus, from the figures (\ref{fig:d41}-\ref{fig:d45}) as well as from the Table-\ref{tab:table1}, it is clear that radiation pressure and P-R drag of both the primaries affect the linear stability range of the problem significantly. The nonzero value of the zero $(\mu_0)$ of the determinant $D_4$ in the Arnold-Moser theorem under non-resonance case, insure the instability of triangular equilibrium points, within the range of stability $0<\mu<\mu_c$.
 
\begin{table}
\caption{Zero $(\mu_0)$ of $D_4$ and critical mass ratio $\mu_c$ at different values of perturbing parameters.}
\label{tab:table1}
\begin{tabular}{@{}|rrrrrr|@{}} 
\hline
$W_1$& $W_1$& $\epsilon_1$&$\epsilon_2$ &$\mu_0$&$\mu_c$\\	
\hline
 0.000&0.0000&0.000&0.00&0.010950&0.0385209\\
 &&&&&\\
 0.005&0.0000&0.000&0.00&0.000305&0.0393447\\
 0.010&0.0000&0.000&0.00&0.000876&0.0401684\\
 0.015&0.0000&0.000&0.00&0.005844&0.0409922\\
 0.020&0.0000&0.000&0.00&0.010970&0.0418159\\
 0.025&0.0000&0.000&0.00&0.016230&0.0426397\\
 &&&&&\\
 0.000&0.0040&0.000&0.00&0.001122&0.0388504\\
 0.000&0.0045&0.000&0.00&0.000440&0.0388916\\
 0.000&0.0050&0.000&0.00&0.000529&0.0389328\\
 0.000&0.0055&0.000&0.00&0.000625&0.0389740\\
 0.000&0.0060&0.000&0.00&0.000727&0.0390152\\
 &&&&&\\
 0.000&0.0000&0.001&0.00&0.014260&0.0385387\\
 0.000&0.0000&0.002&0.00&0.015750&0.0385566\\
 0.000&0.0000&0.003&0.00&0.016580&0.0385744\\
 0.000&0.0000&0.004&0.00&0.017320&0.0385922\\
 0.000&0.0000&0.005&0.00&0.017860&0.0386101\\
 &&&&&\\
 0.000&0.0000&0.000&0.01&0.018040&0.0381642\\
 0.000&0.0000&0.000&0.02&0.015740&0.0378075\\
 0.000&0.0000&0.000&0.03&0.013380&0.0374508\\
 0.000&0.0000&0.000&0.04&0.011130&0.0370941\\
 0.000&0.0000&0.000&0.05&0.008876&0.0367374\\
 &&&&&\\
 0.015&0.0050&0.003&0.031&0.02483&0.0438750\\
\hline
\end{tabular}
\end{table}

\section{Conclusions}
\label{sec:con}
We have considered the photogravitational restricted three body problem in the presence of radiation pressure force and P-R drag of both the massive bodies, which are radiating in nature. Analysis of nonlinear stability of the triangular equilibrium points is performed in non-resonance case using Arnold-Moser theorem under the influence of four perturbing parameters in the form of P-R drags $W_1, \, W_2$ and mass reduction factors $q_1,\,q_2$, of both the primaries. First, we have normalized the Hamiltonian of the problem up to order four using Lie transform method and then Birkhoff normal form of the Hamiltonian constructed, which is necessary to apply the Arnold-Moser theorem in non-resonance case. The determinant $D_4$ of the Arnold-Moser theorem is computed analytically under the consideration of only linear order terms of perturbing parameters, which agree with that of \cite{Deprit1967AJ.....72..173D,Mayer,Kushvah2007Ap&SS.312..279K,kishor2017Ap&SS.362..156K} in the absence of perturbing parameters. To apply the Arnold-Moser theorem in non-resonance case, we have plotted the determinant $D_4$ with respect to the mass parameter $\mu$ within the stability range $0<\mu<\mu_c$. It is observed that in presence as well as in absence of perturbing parameters, there exist a nonzero value of $\mu=\mu_0$ at which $D_4$ vanish (figures (\ref{fig:d41}-\ref{fig:d45})), which insure that triangular equilibrium points are unstable in nonlinear sense.  The effect of perturbing parameters are also analyzed and it is found that on increasing the values of $W_1,\,W_1,\,\epsilon_1$, critical mass ratio $\mu_c$ increases, with the existence of nonzero  $\mu_0$ in each case , whereas on increasing the value of $\epsilon_2$, $\mu_c$ decreases with the existence of nonzero $\mu_0$ (Figure (\ref{fig:muc}) and Table-\ref{tab:table1}). A similar trend is also seen in case of $\mu_{c1}$ and $\mu_{c2}$ (Table-\ref{tab:muc12}). Thus, we conclude that due to radiation pressure and P-R drag of both the primaries, the linear stability range of the problem get changed, significantly. Also, due to existence of nonzero value of the zero $(\mu_0)$ of the determinant $D_4$ in the Arnold-Moser theorem under non-resonance case, within the range of stability $0<\mu<\mu_c$, triangular equilibrium points are unstable in nonlinear sense.  Present analysis is limited up to first order terms of the perturbing parameter, which may be extended to higher order inclusion of the terms. The results obtained can help to analyze the more generalized problem under the influence of other perturbations such as albedo, solar wind drag, Stokes drag etc.

\acknowledgements{We all are thankful to the Inter-University Center for Astronomy and Astrophysics (IUCAA), Pune for providing references through its library and computation facility in addition to local hospitality.  First author is also thankful to UGC, New Delhi for providing financial support through UGC Start-up Research Grant No.-F.30-356/2017(BSR).}

\bibliographystyle{spbasic} 
\bibliography{ref}
\end{document}